\documentstyle[11pt,dunk2001_asp,twoside]{article}
\def\edcomment#1{\iffalse\marginpar{\raggedright\sl#1\/}\else\relax\fi}
\reversemarginpar

\begin{document}
\title{The Infall of Gas onto the Galactic Disk}
\author{Brad K. Gibson}
\affil{Centre for Astrophysics \& Supercomputing, Swinburne University,\\
Mail \#31, P.O. Box 218, Hawthorn, Victoria, Australia, 3122}

\begin{abstract}
Ongoing accretion of low-metallicity gas onto the disk is a natural prediction
of semi-analytical Galactic chemical evolution models.  This 
star formation fuel ameliorates the overproduction of metal-poor 
G- and K-dwarfs in the solar neighbourhood which otherwise plague so-called
``closed-box'' models of Galaxy evolution.  Do High-Velocity Clouds (HVCs)
represent the source of this necessary fuel?  We know that HVCs provide
an important clue as to the processes governing galaxy formation and 
evolution - what is less clear is whether their role lies more closely
aligned with cosmology (as relics of the Local Group's formation) or star
formation (as tidal debris from nearby disrupted dwarfs, or the
waste byproducts of disk supernova-driven winds).  I provide
a summary of recent speculations as to the origins of HVCs, and highlight
several future projects which will lead to a deeper understanding of the
role they play in galaxy evolution.
\end{abstract}

\section{Introduction}

Analytical models of Galactic chemical evolution invariably demand the
existence of infalling (near?) pristine gas onto the disk in order to 
avoid the overproduction of low-metallicity stars - the so-called G-dwarf
problem (Flynn \& Morell 1997).  Larson (1972) first suggested that this 
requisite star formation fuel might be associated with the population of 
High-Velocity Clouds (HVCs) seen moving at anomalous velocities with respect 
to differential Galactic rotation (Wakker \& van Woerden 1997).
Modern state-of-the-art chemical evolution models retain the need for
this infalling fuel (e.g. Chiappini et~al. 1997), although most
do not necessarily target the HVCs as the most likely culprit.
Tosi (1988) suggests that infalling fuel more metal-rich than 
$\ga$0.2\,Z$_\odot$ violates the present-day disk abundance constraints
provided by HII regions, an hypothesis we are exploring with
updated dual-infall (halo+disk phases) models (Chiappini et~al. 2002).

Independent of these Galactic fuel arguments, a natural byproduct
of hierarchical clustering galaxy formation scenarios (such as the
currently favoured $\Lambda$CDM) is that the halo of our Milky Way
should be populated with $\sim$500 satellites (Klypin et~al. 1999), and
accretion of gas should continue (at some level) to the present-day (as in
the aforementioned analytical models).
This prediction is more than an order of magnitude discrepant with that
actually observed ($\sim$30 satellites).  Blitz et~al. (1999) and Braun
\& Burton (1999) have both recently 
revived the classic hypothesis due to Verschuur (1969), suggesting that
HVCs are Local Group interlopers.  This Local Group Infall scenario
is based upon the assumption that the gas we see as an HVC traces an 
underlying (dominant) dark matter halo.

Do HVCs represent the reservoir of star formation fuel predicted 
to exist by analytical and numerical simulations of galaxy
formation?  Are they cosmological relics, waste byproducts related to
tidal disruption of neighbouring dwarf galaxies, or supernova-driven
ejecta from the disk?  Whether they be cosmological, or related to star
formation processes, HVCs represent crucial, yet mysterious, ingredients 
of galaxy formation.  Under the former, the HI component of HVCs would
contribute on the order of 10$^{11}$\,M$_\odot$ to the Local Group, while
the latter\footnote{Assuming typical distances of 4\,kpc, and the
integrated HI flux density from Table~3 of Wakker \& van~Woerden 1991.} 
would correspond to order 10$^{7}$\,M$_\odot$ of HI in the halo.

In what follows, I present a summary of the present-day state-of-affairs in
HVC research, highlighting several intriguing (if sometimes confusing
and/or contradictory) pieces of the puzzle, and avenues of future research.

\section{Weighing the Evidence}

In terms of assessing the role played by HVCs - cosmological or Galactic
structure\footnote{In reality, there is certainly a continuum of HVC
populations; in fact, I would be surprised if there were not (c.f. Blitz 
2001).} - there are numerous discriminants which can be employed to
argue for or against either option.  Unfortunately, virtually all of
these discriminants are purely of an indirect nature in the sense that
unless the property under discussion is highly extreme, its interpretation
is subject to one or more caveats.

Examples of such indirect arguments include:
\begin{itemize}
\item{\it Kinematic Distribution}: While it is tempting to use
the fact that the velocity dispersion of the HVC distribution function
is lower in the Local Group (and Galactic) Standards of Rest than it is in
the Local Standard of Rest (e.g. Blitz et~al. 1999), to argue for
an extragalactic HVC residency, Gibson et~al. (2001a) have shown that this
is argument is specious.
\item{\it Size-Linewidth Relationship}: Combes \& Charmandaris (2000)
note that Galactic halo HVCs adhere to the molecular cloud
size-linewidth relationship, and if the Braun \& Burton (1999) 
Compact HVCs were at 20\,kpc they would also follow this relationship.
While possibly correct, this argument (by itself) is also specious
(Gibson et~al. 2001b).
\item{\it Connection to Local Star Formation}: An unappreciated 
clue to the HVC puzzle is provided by the work of 
Schulman et~al. (1997).  Their VLA analysis of isolated spirals
shows that the incidence of HVC activity is directly related to the
magnitude of underlying star formation in the galaxy's disk.  This
result is based upon only a handful of targets, but is not a prediction of
the Blitz et~al. (1999) model.
\item{\it Local Group Analogs}: To date, extensive HI mapping of nearby
Local Group analogs has not turned up evidence for extragalactic
HVCs in these systems (Zwaan 2001).  Similarly, the statistics of MgII and
Lyman Limit absorbers in the spectra of background QSOs are in disagreement
with the extragalactic HVC scenarios (Charlton et~al. 2000).  By definition,
these other groups are not the Local Group (of course!), but unless one 
argues that the Local Group is somehow privileged, it is difficult to 
escape the obvious implications (c.f. Blitz 2001).
\item{\it Metallicities}: HVC metallicities offer a potentially useful 
discriminant between competing theories for their origin.  Very
low metallicities ($\la$0.01\,Z$_\odot$) would argue against
a recent Galactic disk event (such as a fountain), and support the
Local Group building block scenarios of Blitz et~al. (1999) and
Braun \& Burton (1999).  Conversely, very high metallicities
($\ga$Z$_\odot$) would be consistent with a Galactic fountain, but 
incompatible with the infalling star formation fuel hypothesis
(recall the metalllicity argument of Tosi 1988).  It should not be
surprising to read that the interpretation of intermediate 
metallicities would be somewhat more contentious!

To date, there are only a few HVCs for which accurate metallicities exist - 
none show evidence for metallicities below 0.1\,Z$_\odot$, but neither
do any show evidence for metallicities above 0.5\,Z$_\odot$ (Gibson et~al.
2001b,c; Wakker 2001).  In fact, only one HVC has been claimed to have
a metallicity $\la$0.2\,Z$_\odot$ - Complex~C (Wakker et~al. 1999).
Gibson et~al. (2001c) have shown though that Complex~C is not the simple
0.1\,Z$_\odot$ cloud that Wakker et~al. thought it was.  In fact, along
one of the sightlines (Mrk~817), the sulfur and oxygen abundances
are clearly in excess of $\ga$0.2\,Z$_\odot$.  Recall, that both elements
are depleted only lightly onto dust and the ionisation corrections for
both OI and SII are (effectively) impervious to ionisation corrections.
In other words, the metallicity determinations along these sightlines
are robust (modulo HI column density normalisation, derived from lower
spatial resolution 21cm data).  Adopting the most recent solar system
abundance normalisation increases the inferred oxygen abundance for
the Mrk~817 sightline to 0.31\,Z$_\odot$ (Gibson et~al. 2001c).

Taken together, it is tempting to suggest that virtually all HVCs have
metallicities in the narrow range 0.2$\la$Z$_\odot$$\la$0.3. At face
value, this is the metallicity one would expect a priori for 
tidally-disrupted debris from the Large or Small Magellanic Clouds
(Gibson et~al. 2000) \it or \rm from disk gas which originated at
Galactocentric distances in excess of 
$\sim$1.5\,r$_\odot$ (Gibson et~al. 2001b; Figure~2b).  \it If \rm
disk gas could be diluted in metallicity 
by a factor of $\sim$2 during its putative
passage through the halo even inner disk gas could be the ultimate source
of HVCs.

Interpreting the known HVC metallicities is difficult.  While we can 
apparently rule out simple, undiluted, Galactic fountains, we can also
rule out the failed dwarf/Local Group formation remnant scenario (recall
Figure~1 of Gibson et~al. 2001b) as all HVCs appear to be at least
10$-$30$\times$ more metal-rich than the Local Group dwarf spheroidals.

\item{\it Connection to Galactic Gas}: A danger with using digital tables
of HVC properties (such as Wakker \& van Woerden 1991 and Putman et~al 2002)
is that the user may not appreciate that arbitrary cuts in $v_{\rm LSR}$ may
mask sinuous connections to low-velocity Galactic gas.  As Cohen (1981)
and Putman \& Gibson (1999) have shown, such connections are not uncommon
and clearly show that some HVCs are most definitely of a Galactic nature.
Of course, this argument is not valid for \it all \rm HVCs.
\item{\it Alternatives to $\Lambda$CDM}: Perhaps the simplest way in which
to eliminate any discrepancy between the predictions for $\Lambda$CDM
satellite counts versus those observed is not to associate HVCs with the
$\Lambda$CDM halos, but to find a way to hide the halo baryons or eliminate
the halos altogether.  The former might include efficient feedback
and/or photoionisation at high-redshift (Bullock et~al. 2000;
Chiu et~al. 2001; Somerville 2001), while the latter might include
variants to $\Lambda$CDM itself, such as Warm Darm Matter
(Knebe et~al. 2001), Self-Interacting Dark Matter
(Dav\'e et~al. 2001), or Tilted Cold Dark Matter (Bullock 2001).
The theorist in me enjoys the fevered investigations into CDM alternatives,
while the observer remains ever skeptical ...
\end{itemize}

There is really only one {\it pure} 
discriminant between these extra-Galactic and
Galactic scenarios - knowledge of the distance to the population of
HVCs.  Unfortunately, the distances to all but a few HVCs are unknown.
Doubly unfortunate is the fact that there is really only one way to 
derive an unequivocal HVC distance - the detection of the cloud in
absorption against a background halo star of known distance.
\begin{itemize}
\item{\it Absorption Line Distances}: Seen in absorption against a range
of blue halo stars, five HVCs clearly reside in the halo (Gibson et~al.
2001a; Table~1).  Of course, this technique can \it only \rm be
applied to clouds in the halo - even if a given HVC did reside outside the
halo, there would be no sufficiently bright background star to use as the
probe of the intervening cloud.  Clearly these five HVCs are inconsistent
with the Local Group Infall models of Blitz et~al. (1999) and Braun \&
Burton (1999).
\end{itemize}

Because of the difficulty in applying the absorption line distance technique,
a number of indirect distance determinants have been proposed, including:
\begin{itemize}
\item{\it H$\alpha$ Distances}: Under the (reasonable)
assumption that $\ga$1\% of the disk's ionising escape into the halo, coupled
with a model describing the photon sources' distribution, as well as
some knowledge of the covering
fraction, topology, and line-of-sight orientation of a given HI screen, the
measured H$\alpha$ emission measure from the screen can be inverted to
provide a distance.  The halo ionising radiation field model of 
Bland-Hawthorn et~al. (1999,2001) is generally adopted; practical
applications of the technique have been demonstrated by
Bland-Hawthorn et~al. (1998), Tufte et~al. (1998), and Weiner et~al. (2001).
The technique has received a great deal of attention due to its potential
application across a large fraction of the HVC population; that said, it
has also been one of the more contentious issues in the field!  Many of
its shortcomings appear to have been rectified by the recent inclusion
of spiral arms into the underlying disk model (Bland-Hawthorn et~al. 2001).
Weiner et~al. (2001) suggest that the extant data is inconsistent with the
Blitz et~al. (1999) Local Group Infall model.
\item{\it Head-Tail Substructure}: $\sim$20\% of Compact HVCs show distinct
head-tail structure, suggestive of interaction with an external medium.
Quilis \& Moore (2001) show that an ambient density in excess of
$\sim$10$^{-4}$\,cm$^{-3}$ is required to shape these clouds.  Such
densities are consistent with an upper halo location, but \it not \rm
an intergalactic one.  That said, this argument only applies to a subset of 
HVCs.
\item{\it Pressure Arguments}: Burton et~al. (2001) employ 
thermal pressure arguments applied to Compact HVCs to claim that HVCs
lie at distances of 400$\pm$280\,kpc.  As pointed out by
Amiel Sternberg (cited in Gibson et~al. 2001b), 
this result should strictly be interpreted as an
upper limit (and not an equality), since Burton et~al. only adopt the
minimum pressure necessary to maintain the observed core/halo interface
(as opposed to the range allowed by a multiphased mixture).
\item{\it Cohen Stream}: The Cohen Stream (HVC~165$-$43$-$120),
part of the Anti-Centre High-Velocity complex mapped in HI by Cohen (1981),
spans 25$^\circ$ on the sky.  This $-$120\,km/s feature is not seen in
H$\alpha$, but it does trace a parallel HI filament at $-$13\,km/s, and so
must be at a distance $\la$300\,pc.
\item{\it Search for Stars in HVCs}: The search for stars within HVCs
(and especially Compact HVCs) has begun in earnest.  Programs are underway
at LCO/KPNO (Grebel et~al. 2000), the Sloan Digital Sky Survey (Willman 
et~al. 2002), and via the use of archival POSS-II plates (Simon \& Blitz
2002, in preparation) - to date, none of these studies have uncovered a
stellar population associated with any HVC.  A primary motivation for
this work is the search for an excess of potential red giant branch tip
stars (useful distance diagnostics) and (perhaps) even RR~Lyrae.
\end{itemize}

\section{The Future}

An enormous effort is underway to map stellar sub-structure in
the halo, and use this information to reconstruct the detailed physics
governing the formation of the Milky Way (e.g. Ibata et~al. 2001; 
Morrison et~al. 2000;
Helmi et~al. 1999; Willman et~al. 2002).  
It is crucial to remember that the halo is populated
by not only stars, but a not insignificant reservoir of gas.
This gas distribution needs to be painstakingly mapped, its origin(s)
assessed, and the relevant physics explored - whether this physics is
more aligned with cosmology or star formation is yet to be definitively
demonstrated, but does \it not \rm change the fact that HVCs are
important clues to galaxy formation \it and/or \rm galaxy evolution!  
\it Halo gas tomography \rm is the most pressing area of
HVC research.  

With Mike Bessell, Tim Beers, Norbert Christlieb, and 
Joss Bland-Hawthorn, we have embarked on a long-term program using the
6dF Facility at the United Kingdom Schmidt Telescope to identify large
numbers of suitable (distant) 
halo probes aligned with (potenially) foreground HVCs.  Full three-dimensional
mapping of halo HVCs is our ultimate goal.  In conjunction with the
stellar tomography provided by satellites such as FAME, DIVA, and (especially)
GAIA, halo gas tomography will provide the ingredients necessary to undertake
full baryonic reconstructions of the halo's formation.
These tests of \it near-field cosmology \rm are poised to compete on an
equal level with \it far-field cosmology \rm over the coming decade.

High-resolution N-body and hydrodynamical simulations of both the interaction
of HVCs with the Galaxy, and the Local Group's formation, must continue
for the foreseeable future.  As Quilis \& Moore (2001) demonstrate, there
is much yet to be learned.  With Daisuke Kawata, for example, we have been
attempting to simulate the spatial and kinematic properties of Complex~C
using satellite perturbers (as a sort of ``straw man'' model) interacting
with the Outer Arm of the Milky Way's disk.  We have
actually been (reasonably!) successful in doing so, generating 
one realisation in which Complex~C gas originated from within the
disk (at $\sim$2\,$r_\odot$), but from the opposite side of the Galaxy.
Gas was drawn from the disk by the close passage of the perturber and
``pulled'' over the Galaxy in its wake, to find itself now infalling onto
the disk in the vicinity of the Outer Arm at $\ell$$\sim$90$^\circ$.
The spatial, kinematical, and metallicity properties of Complex~C are
all recovered!\footnote{At the expense of doing some substantial damage to
the gas disk in the inner regions of the Milky Way, but we'll sweep that
under the rug for another day (Gibson \& Kawata 2001, in preparation)!}
In addition, we are pursuing very high mass resolution sticky-particle
simulations of the Local Group's formation with Kenji Bekki.

The availability of STIS and (soon) COS on HST, as well as FUSE, means that
a concerted effort needs to be made in continuing to obtain metallicities
and ionisation conditions of HVCs (and IVCs) using background QSO probes.
Time Allocation Committees need to be made aware that not all HVCs are the
same, and just because one or two clouds have well-determined 
metallicities does not mean that therefore all clouds have now been sampled!
We have been successful in obtaining FUSE Cycle~2 time to study the anti-Local
Group barycentre HVC~246$+$39$+$125, and have recently been awarded HST/STIS
Cycle~11 time to study the Compact HVC~225$-$83$-$197 (to supplement our
extant FUSE data on this sightline).  Cloud abundance patterns remain
a powerful tool for disentangling different origins scenarios (and
determining the nucleosynthetic pollution history - e.g. have these
clouds seen pollution from Type~II or Type~Ia supernovae?).  Despite these
successes, it is most definitely a challenge to convince TACs that HVCs
are important clues to the formation mechanisms of galaxies!

Further searches for HVCs in nearby Local Group analogues must continue,
as the initial results from Zwaan (2001) have been so intriguing.
With D.J. Pisano, David Barnes, and Lister Staveley-Smith, we are using
the narrowband filters with the mulitbeam facility at Parkes to Nyquist
sample $\sim$1\,Mpc$^2$ around Local Group ``clones'' to an HI mass
sensitivity of 3$\times$10$^6$\,M$_\odot$ (5$\sigma$).  With B\"arbel
Koribalski, we are surveying (to a comparable sensitivity limit)
an $\sim$100\,sq.deg region of the
Supergalactic Plane in the vicinity of the HVC (protogalaxy?) discovered
by Kilborn et~al. (2000), in order to search for low-level extended emission.

On an even more
ambitious level, with Frank Briggs, Martin Zwaan, David Barnes, et~al., we
have been awarded 14 nights at Parkes as a pilot study to what we call
HIPARK - a parked, transit-style, Parkes, survey.  Ultimately, our goal is 
to drift-scan the same ($\sim$100\,sq.deg) 
strip of sky over a continuous three-month
period, down to a 6$\sigma$ limiting HI mass of 3$\times$10$^4$\,M$_\odot$
(at 1\,Mpc)!\footnote{Or 3\,M$_\odot$ at 10\,kpc!}  

Finally, a concerted effort must be made to either find stars associated
with HVCs (they are enriched in metals, and if they are at Local Group-like
distances, they almost certainly had to acquire those metals internally 
through star formation) or definitively refute their existence (which
would argue for them being local, and enriched from their parent galaxy,
whether that was the Milky Way, LMC, SMC, or some other disrupted dwarf).
Many Time Allocation Committees have recognised the value of these searches,
but that is most definitely not a universal sentiment!  Deep wide-field
imaging of a statistically significant number of Compact HVCs (and
neighbouring blank sky) should be undertaken at 4-8m class facilities
as soon as possible.

\acknowledgments
First, let me state my appreciation to Ken for all the little things
he has done to help my career.  Ken will leave
an unprecedented legacy in not only science, but in the mentorship role
he played for Stromlo students and postdocs (myself included)
alike.  He was one of the primary
reasons I originally came to Australia in '95, and 
one of the reasons I returned again (permanently) in '00.
Let me also take this opportunity to thank Mary Putman; I was fortunate
in the extreme to have had such an excellent PhD student first time around!
Thanks also to my various HVC collaborators, including
Yeshe Fenner, Daisuke Kawata, Sarah Maddison, Cristina Chiappini, 
Francesca Matteucci, B\"arbel Korbalski, Lister Staveley-Smith, Frank Briggs,
Mary Putman, Joss Bland-Hawthorn, Phil Maloney, D.J. Pisano, Ken Freeman,
David Barnes, Mark Giroux, Kenji Bekki, Martin Zwaan, Bart Wakker,
John Stocke, Mike Shull, Ken Sembach, Mike Bessell, Tim Beers, and
Norbert Christlieb.
The financial support of both the Australian Research Council and the
Victorian Partnership for Advanced Computing is acknowledged.  Finally,
my thanks to Gary Da Costa and the Organising Committee for providing
travel assistance, and seeing to the seamless operation of a wonderful
conference.

\end{document}